\documentclass{article}
\usepackage[margin=1in]{geometry}

\usepackage{pgfplots}
\pgfplotsset{compat=1.15}

\usepackage{listings}
\usepackage{multicol}
\usepackage{color}
\usepackage{amsmath}
\usepackage{textcomp}
\usepackage{amssymb}
\usepackage{amsthm}
\usepackage{mdframed}
\usepackage{algorithm}
\usepackage[backend=biber]{biblatex}
\usepackage[noend]{algpseudocode}
\usepackage{subcaption}

\makeatletter
\def\BState{\State\hskip-\ALG@thistlm}
\makeatother

\newcommand\todo[1]{\textcolor{gray}{TODO: #1}}
\renewcommand\todo[1]{}   

\definecolor{dkgreen}{rgb}{0,0.6,0}
\definecolor{gray}{rgb}{0.5,0.5,0.5}
\definecolor{mauve}{rgb}{0.58,0,0.82}
\let\vec\mathbf
\DeclareMathOperator{\sym}{sym}
\DeclareMathOperator{\tr}{tr}

\theoremstyle{definition}
\newtheorem{sidenote}{Side-note}[section]

\usepackage[final]{hyperref} 

\usepackage{footnotebackref}
\usepackage{cleveref}

\hypersetup{
    colorlinks=true,       
    linkcolor=blue,        
    citecolor=blue,        
    filecolor=magenta,     
    urlcolor=blue
}

\title{Inferring ice sheet damage models from limited observations
using CRIKit: the Constitutive Relation Inference Toolkit}

\author{Grant Bruer, Tobin Isaac}

\begin{document}

\maketitle

\abstract{%
  We examine the prospect of learning ice sheet damage models from
  observational data.  Our approach, implemented in CRIKit (the Constitutive
  Relation Inference Toolkit), is to model the material time derivative of
  damage as a frame-invariant neural network, and to optimize the parameters
  of the model from simulations of the flow of an ice dome. 
  Using the model of Albrecht and Levermann as the ground truth to generate
  synthetic observations, we measure the difference of optimized neural
  network models from that model to try to understand how well
  this process generates models that can then transfer to other ice sheet
  simulations.

  The use of so-called "deep-learning" models for constitutive equations,
  equations of state, sub-grid-scale processes, and other pointwise relations
  that appear in systems of PDEs has been successful in other disciplines, yet
  our inference setting has some confounding factors.
  The first is the type of observations that are available: we compare
  the quality of the inferred models when the loss of the numerical
  simulations includes observation misfits throughout the ice, which is
  unobtainable in real settings, to losses that include only combinations
  of surface and borehole observations.
  The second confounding factor is the evolution of damage in an ice sheet,
  which is advection dominated.  The non-local effect of
  perturbations in a damage models results in loss functions that have both many
  local minima and many parameter configurations for which the system is
  unsolvable.

  Our experience suggests that basic neural networks have several deficiencies
  that affect the quality of the optimized models.  We suggest several
  approaches to incorporating additional inductive biases into neural networks
  which may lead to better performance in future work.
}%

\section{Introduction}

When modeling real-world phenomena with partial differential equations
(PDEs), behavior is governed by a combination of principles that are
considered inviolable (such as conservation principles) and
other processes that are more heuristic.  The latter include equations
of state and constitutive relations that have some formal constraints --
such as frame invariance, monotonicity, or convexity requirements --
but not enough to define them completely, and so require some data to additionally constrain them.  Classically, this data could come from, for example, laboratory experiments, which generated data to fit simple parametric models of the processes.  Many classical models of equations of state or constitutive relations consist of just one parameter, such as the thermal diffusivity of a material, but often these classical models are too simple,
and more complex parametric models are required.

Let $d = f(\sigma)$ be a description of one such process: we think of it as being
a relationship that acts pointwise on some measurable state of the system, 
$d(x) = f(\sigma(x))$.
If $f$ has some free parameters, $f(\sigma) = f(\sigma;\theta)$,
then the process of designing and constraining such parametric models of constitutive relations can be thought of, in the language of machine learning,  as a form of supervised learning, where data collected from multiple experiments defines a loss $\ell$ on the model parameters to be minimized.  Simple laboratory experiments are designed to have states that are as spatially constant possible, $\sigma(x) \equiv \sigma$, so that each individual experiment $i$ roughly gives us a training pair $(\sigma_i, d_i)$ of direct inputs and outputs of the model~$f$.  Thus with enough experiments, the optimal parameters minimize the sum of \emph{separable} contributions to the loss,
\begin{equation}\label{eq:separable}
\theta^* = \arg\min_\theta \sum_i \ell(f(\sigma_i),d_i).
\end{equation}

Not every constitutive relation of interest, however, can be studied in a laboratory setting.  This is particularly true of relations that appear in geophysics, where the material of some physical feature, such as a landform or glacier, can only be studied by much less controlled experiments, and often indirectly.  The mathematical approach to finding the optimal parameters of $f$ in this case must include some model $A$ of PDEs governing the system, in which $f$ appears as a closure term, as well as some model $B$ of how the state of the system could have generated the observed data.  That data is often not separated into uncorrelated observations, and instead appears as one variable $d$, so that the optimal parameters of $f$, in contrast to \cref{eq:separable}, is a \emph{monolithic} PDE-constrained optimization problem,
\begin{equation}\label{eq:pdeconst}
\theta^* = \arg\min_{\sigma,\theta} \ell(B(\sigma), d)\quad \text{subject to}\  A(\sigma;f) = 0.
\end{equation}

The efficiency with which modern machine-learning tools can train expressive parametric models such as deep neural networks from large volumes of data has led to recent successes in using deep neural networks as constitutive relations
(or similar pointwise processes) in PDE simulations.  Examples of these successes include the training of subgrid models of turbulence in the Navier-Stokes equations \cite{ling_kurzawski_templeton_2016,gamahara2017searching}.  These networks are trained from a loss function comparing the solution fields of PDE simulations to ground-truth data (although in both of those works the PDE solutions are not re-evaluated with the optimized pointwise functions, so they stop short of implementing full PDE-constrained optimization as described in 
\cref{eq:pdeconst}).  This suggests to geophysicists that the same approach may work for their problems in constitutive relation modeling.

Yet, even within the framework of PDE-constrained optimization, the volume of relative independence of components of the data $d$ can vary greatly.
In the Navier-Stokes example cited, large volumes of DNS simulations can be used to generate data for loss function on subgrid closure models.
In the modeling of actual glaciers and ice sheets, however, the observation
process $B$ in the PDE-constrained optimization problem must reflect the actual mechanisms at hand, such as satellite observations and borehole samples,
which are both sparser than simulated DNS data and appear only on low-dimensional surfaces of the ice, and not throughout.

This work investigates the possibility of using neural networks, which are only weakly constrained in form, to approximate the constitutive relationships that appear in ice sheet modeling.  In particular, we seek to understand how accurately they can be trained to generate true constitutive relations, which
can be expected to behave accurately if transferred to a different ice geometry
or forcing condition than the ones under which they were trained.

The use of neural networks as part of modeling PDEs has been the focus of much recent research.  Including their use as surrogate models \cite{SUN2020112732}, deep neural networks have been used as discretization-free PDE evolution models in Physics-Informed Neural Networks \cite{raissi2019physics}.  In contrast, a properly trained constitutive relation should not encode any of the geometric aspects of the training data, since it is evaluated at each quadrature point and takes only local data as inputs.  We also contrast our approach with sparse identification of nonlinear dynamics (SINDy) \cite{brunton2016discovering}, in which governing equations for nonlinear dynamics are trained from time series data.  For one, SINDy solutions are optimized from supervised experiments, where the direct outputs of the network are available.  SINDy also relies on a predetermined library of nonlinear operations, of which it tries to ascribe the behavior to only a sparse few.  A SINDy-trained function thus has some notion of explainability of the result: in contrast, the trained neural networks that appear later in our work identify no simple functional form.

Large scale ice sheet models generally use continuum representations of all of the processes within the ice sheet, including the dynamic variables such as velocity and pressure, but also hydrology and mechanical properties of the ice.  Glacier and marine ice is known to be highly heterogeneous, with the effective rheology of the ice being affected by things like the climatic conditions under which it first fell as snow.  Of particular interest, and the focus of this work, is the way that the ice fabric evolves under different ambient conditions of stress, strain, temperature, etc.  The conditions affect grain alignment and dislocation of molecules between grains, and as a result the fabric becomes not just stronger or weaker, but develops anisotropic responses to strain.  These processes are too complex to represent directly in a large-scale simulation, and so continuum upscaling is required.  This continuum representation takes the form of a constitutive relation, and we refer to it as the ``damage model'' used in the ice evolution equations.

There is no first-principles explanation that determines the functional form that a damage model must take.  In \cref{sec:model_survey} we survey damage models that have appeared in the literature recently.  These models are parametric, with parameters that must be constrained by observations.  Because these models are derived by hand, they typically have a very simple functional form: only as many terms are present as are necessary to preserve some type of qualitative behavior, such as a monotonic increase in the rate of damage under increasing shear stress.

How much do the functional forms of these hand-derived damage models bias the behavior of the ice?  Are all damage models that can reproduce existing observations ``close'' to the hand-derived ones in some sense, or are there
equally good damage models that are qualitatively different?  This work addresses these questions by formulating damage models as general neural networks, which are trained to match observations that resemble field measurements.  Once we have done this, we investigate whether these trained networks converge to damage models that are close to the hand-derived ones.

\todo{cite "External Operators in Firedrake" because their work is very relevant to ours}

\section{Background}

\subsection{Ice flow simulation}

In this work we treat ice as a creeping, incompressible material, so the PDEs governing the gravity-driven steady-state system consist of the force balance equation
and the incompressibility condition,

\begin{align}
- \nabla \cdot \underbrace{(\tau - p I)}_{\sigma} &= \rho \vec{g} \label{eq:force-balance-pde} \\
\nabla \cdot \vec{u} &= 0 \label{eq:incompressibility-pde}
\end{align}

On the left hand side, $\vec{u}$ is the unknown velocity, $p$ is the unknown pressure, $\tau$ is the deviatoric part of the stress tensor $\sigma$, and $I$ is the
identity tensor.  On the right hand side, $\rho$ is the density of the ice and $\vec{g}$ is
the gravitational acceleration vector.
A complete ice sheet or glacier model must also include
temperature or enthalpy as a indeterminate state variable, but in this work we consider the isothermal approximation, and will treat the temperature $T$ as a constant.

To form a well-posed problem, these equations must be augmented with boundary conditions and with a stress closure relating the deviatoric stress $\tau$ to the local state of the ice.
In the experiments we conduct in this work in \cref{sec:experiment}
we use standard boundary conditions and will discuss them further there.

A numerical discretization of \cref{eq:force-balance-pde,eq:incompressibility-pde} defines a nonlinear systems of equations, where
the stress closure must be evaluated at multiple points in the ice domain: at lattice points in a finite difference discretization, at cell centroids in a finite volume discretization, or at quadrature points in a finite element discretization.  In any of those settings, the discretization scheme determines numerical approximations to the spatial gradients and/or higher derivatives of the state variables, such as the strain-rate tensor
$\dot{\epsilon}(\vec{u}) = \frac 1 2 (\nabla \vec{u} + \nabla \vec{u}^T)$, which appear as terms in the stress closure model.

In this work we consider models of $\tau$ that are independent of the pressure $p$, but which can depend on the temperature $T$, velocity $\vec{u}$, and other state variables not yet introduced, as described in the following section.


\subsection{Stress-strain closures in ice}

Ice sheet simulations can vary in complexity from shallow-ice or shallow-shelf approximations  to full three-dimensional Stokes simulations. They may take into account various effects such as temperature evolution, firn compressibility, bed topology, bed sliding, and acceleration. However, no matter what simplifying assumptions are made in a simulation, it must include a stress-strain closure that captures how the ice responds to an applied stress.


The constitutive relations typically used in ice sheet modeling are based on Glen's flow law \cite{glen1958flow}, which postulates that the second invariant of the strain rate is proportional to the cube of the second invariant of the stress: $\dot \epsilon_{eff} \propto \tau_{eff}^n$, usually with $n\approx 3$. This law is an empirically-derived approximation that is valid in a wide range of creeping ice flows but is insufficient in many cases. For example, the relation predicts the viscosity goes to infinity as the stress goes to zero, but in reality, the viscosity is finitely bound, indicating a linear relation at low stresses instead of a cubic relation \cite{pettit2003ice}.

The constant of proportionality in Glen's flow law is typically taken to depend on temperature as well as a spatially-varying non-dimensional ``enhancement factor" that attempts to capture any deviations from the original Glen law for a specific ice sheet, such as impurities, grain sizes, and crystal orientations. The enhancement factor is determined empirically based on simulations or measurements of ice \cite{ma2010enhancement,treverrow_budd_jacka_warner_2012} and is not necessarily consistent with any known ice mechanisms.

Crystal orientation is known to depend on strain history, leading to ice becoming more shearable closer to the bed where it is older. The evolution of the crystal orientation over time can be modeled \cite{gillet2006flow} and can then be used in an anisotropic version of Glen's flow law or to compute the scalar enhancement factor in the isotropic version.

An isotropic relation cannot describe every flow field, but it is often a useful approximation. In that regard, a scalar field with a time evolution can directly be used to capture anisotropic weakening. These models that include a time evolution across strain history have the benefit of wider generality. That is, an empirical enhancement field applies only to a specific ice domain, but with a relationship describing the evolution of that enhancement field, the same equations can be applied to a new system or new geometry simply by applying that relationship within the new system.

\subsection{Existing damage models}\label{sec:model_survey}

If we limit the evolution of the enhancement factor at a location in the ice sheet
to depend only on the local conditions at that point and to be independent of arbitrary parameterizations like PDE discretization or the coordinate system used, we inevitably arrive at something better described as a ``damage model'', in which the ice fabric strengthens or weakens based on the local strain, stress, temperature, etc.
Indeed, there are several existing models of more complex effects than a
constant enhancement factor field, ranging from a constant orientation field to
a history dependent field.

Anisotropic simulations are more computationally expensive compared to isotropic
simulations. In many cases, the anisotropy can be captured with a scalar
parameter that enhances the isotropic flow. Scalar models can't capture every
anisotropic effect, but are often useful in real-world regimes.

\textcite{graham2018implementing} built the scalar model ESTAR that takes into
account anisotropic effects. Their model expresses the enhancement factor as a
function of the compressive stress and shear stress, using the relative fraction
to vary the enhancement factor between a compressive enhancement factor and a
shear enhancement factor.
\textcite{albrecht2014fracture} designed a history-dependent scalar damage
model, which advects with the ice.

The General Orthotropic Linear Flow law (GOLF) proposed by
\textcite{gillet-chaulet_2005_anisotropic-flow-law} models the anisotropy with
an orientation distribution field (ODF) with tabulated viscosity parameters (6
at each grid point, computed with a micro-macro model) to capture the
anisotropy. This original model was not history dependent, but
\textcite{gillet2006flow} have extended their model to advect the ODF with the
ice.
\textcite{ma2010enhancement} modified this relation to a nonlinear Glenn flow
form.

We summarize these models in \cref{tbl:enhanced-models}.

\begin{table}
\centering
\begin{tabular}{c|c|c|c}
    Model &
    Citation &
    Functional Form &
    History-dependent \\ \\
Glen Flow & Standard practice & constant field & \\
ESTAR & \textcite{graham2018implementing} & scalar function of strain rate & \\
Damage & \textcite{albrecht2014fracture} & advected scalar &  \checkmark \\
Damage2 & \textcite{borstad2016constitutive} & scalar function of strain rate &  \checkmark \\
GOLF & \textcite{gillet-chaulet_2005_anisotropic-flow-law} & orientation tensor & \\
Advected GOLF & \textcite{gillet2006flow} & advected orientation tensor &  \checkmark \\
Advected Nonlinear GOLF & \textcite{ma2010enhancement} & advected orientation tensor &  \checkmark \\
\end{tabular}
\caption{Overview of ice sheet models that take anisotropic effects into account.} \label{tbl:enhanced-models}
\end{table}

\section{CRIKit approach}





\begin{figure}
\centering
\includegraphics[width=\textwidth]{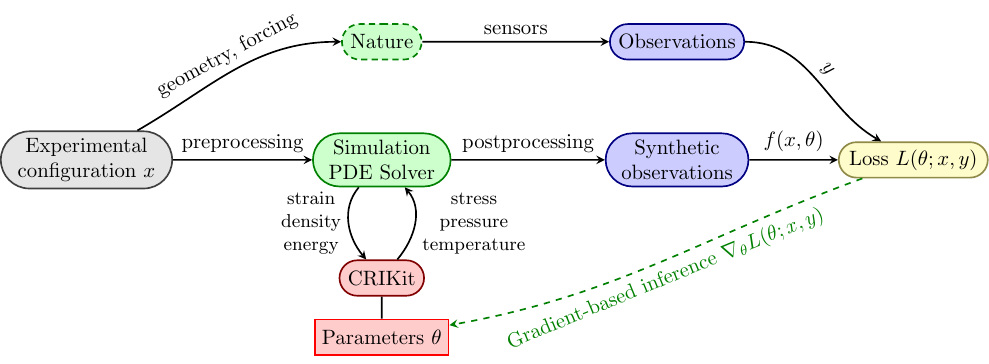}
\caption{A diagram of the CRIKit approach. CRIKit inserts a parametric CR into a simulation so that synthetic observations can be compared to real observations in a loss function.
}\label{fig:crikit_diagram}
\end{figure}

The Constitutive Relations and Inference Toolkit\footnote{\url{https://crikit.science}} (CRIKit) is a software framework designed to construct and train novel constitutive relations that appear in systems of equations governing physical systems.
CR design with CRIKit can be described by four software components.
First, a \emph{CR} expresses the relation between physical quantities; this is independent of any mesh or geometric data. 
Second, an \emph{experiment} inserts the CR into a simulation of a physical system and runs the simulation, e.g., by solving a PDE with the finite-element method, which requires evaluating the CR at quadrature points.
Third, an \emph{observer} simulates measurements of the system  so that these simulated observations can be directly compared to real observations.
Finally, a \emph{loss} function computes a scalar measure of the error between the simulated observations and the real observations.
We show a diagram of the interaction between these components in \cref{fig:crikit_diagram}.

When these components are implemented for a physical system with a parametric CR, the parameters of the CR can be optimized to minimize the loss using standard optimization techniques.
For a first-order optimization method, each step of the optimization requires computing the loss and the loss gradient.
Evaluation of the loss comprises the evaluation of the CR with its current parameters within the modeling of the experiment (typically by solving the discretized PDE), followed by the observation process and scalar error measure.
The evaluation of the gradient of this quantity is an application of backwards-mode chain-rule differentiation, which for the experiment entails the solution of an adjoint model to the original experiment (see for example \cite{mitusch2019dolfin}), and for the CR entails the typical neural network method of backpropagation \cite{kelley1960gradient,werbos1982applications}.

\begin{figure}
    \begin{align*}
    \sum_{i=1}^m \frac{1}{m} |f(\theta)u_i - f_i|^2, &&
    \frac{1}{2}
    \left\|
    \mathbf{d} - 
    \mathbf{B}
    \left\{
    \mathbf{C}^T
    \begin{bmatrix}
    w_1 f(\theta) & & &  \\
    & w_2 f(\theta) & &  \\
    & & \ddots &  \\
    & & & w_m f(\theta) \\
    \end{bmatrix}
    \mathbf{C}
    \right\}^{-1}
    \mathbf{b}
    \right\|^2
    \end{align*}
    \caption{%
      A schematic comparison of separable supervised learning (left)
      and PDE-constrained optimization (right). 
  }
    \label{fig:pdevsml}
\end{figure}

\subsection{Constraining the CR}

We give a schematic representation of the way that PDE-constrained optimization differs from traditional training of neural networks in \cref{fig:pdevsml}. In this simplified example, a parameterized relationship $f(\theta)$ that is linear in the state variables $u$ is being inferred from data. In separable supervised learning, the difference between the model $f$ and expected outputs $f_i$
is evaluated at a predetermined collection of training pairs and averaged.  In PDE-constrained optimization (here we take the simplified example of an linear implicit PDE), there are no predetermined inputs $u_i$: the state variables are determined as part of solving the linear system.  The relation $f$ is evaluated at predetermined quadrature points (represented by the diagonal matrix with quadrature weights $w_i$), but the state variables are not directly modeled at the quadrature points: a finite element interpolation matrix $\mathbf{C}$ maps the degrees of freedom of the discretized systems to those quadrature points.  Each quadrature point combines contributions from multiple degrees of freedom, and the effects of each evaluation of $f$ are further mixed by the inversion of the resulting system matrix,
so that each solution variable $u_j$ is sensitive to contributions from each quadrature point, $w_i f(\theta)$.  Furthermore, the observation operator $\mathbf{B}$ maps the state solution $u$ to observations that are of a completely different class than the direct outputs of the relation $f$.

Physical CRs must be independent of discretization choices, such as the mesh and coordinate system.
If we change the coordinate system of the inputs to a simulation, the outputs of the simulation must change accordingly.
This means a scalar quantity remains unchanged, a vector quantity rotates according to the coordinate change, and higher-order tensors following higher-order coordinate transformation rules.
This coordinate-system invariance can be built into the functional form of the CR by using an invariant representation described by \textcite{wineman1964material}.

In Wineman-Pipkin form, an arbitrary invariant function is represented by a set of arbitrary scalar functions that act as coefficients for a linear combination of known equivariant tensors.
The equivariant tensors automatically capture any coordinate transformation.  
The scalar functions must not change under coordinate transformation, and therefore, must be functions of scalar invariants of the input tensors.

Let $P$ be a tuple of input tensors, $G(P)$ be the tuple of form invariants,
$J(P)$ be the tuple of scalar invariants, and $c(J)$ be the form coefficients. Then, for any invariant tensor function $f(P)$, we can write it in the form
\begin{align}
f(P) = \sum c_i(J(P)) \,\, G_i(P),
\end{align}
where $c_i(J)$ are arbitrary scalar functions.
Although all CRs can be put into this form, they may have different structures. Some models may be expressed as an ODE while others may be expressed as a DAE.
In \cref{app:wineman-pipkin}, we show how some of the relations listed in \cref{tbl:enhanced-models} are expressed in this form.

In this work, we rephrase the problem of determining a constitutive relation $f(P)$ to the problem of determining its invariant function $c(J)$. 
This reduces the input and output space of the function.

\begin{lstlisting}[float, caption={Running experiment, observer, and loss.}, numbers=left, stepnumber=1, label={lst:opt-top-level},escapeinside={(*@}{@*)}]
from crikit import CR

cr = CR(...) # construct a constitutive relation (*@\label{line:top-level-cr}@*)

def run(cr): # numerical experiment incorporating the CR
    from fenics_adjoint import *
    F = ... * dx  # adjoint-enabled variational form with
                  # dummy variables for cr inputs and outputs
    F_with_cr = assemble_with_cr(F, cr, ...) # insert cr into the variational form,
                                             # so it will be called when
                                             # evaluating or assembling the form
    w_pred = solve(NonlinearProblem(F_with_cr)) # compute the predicted state
    return w_pred
    
def observer(w_pred): # model the observation process that generated the data
    y_pred = ...      # (crikit does not play a role in this step)
    return y_pred
    
def loss(y_pred, y_true): # define the loss between a prediction and data
    loss = ...            # (crikit does not play a role in this step)
    return loss

# these operations are taped by pyadjoint
w_pred = run(cr)
y_pred = observer(w_pred)
err = loss(y_pred, y_true)

from pyadjoint import Control, ReducedFunctional, minimize
Jhat = ReducedFunctional(err, [Control(p) for p in cr._params]) (*@\todo{\texttt{\_params -> params?}}@*)
optimized_cr_params = minimize(Jhat)

# use the optimized parameters
cr.set_params(optimized_cr_params)
\end{lstlisting}

\begin{lstlisting}[float, caption={Setting up a parametric invariant power-law CR for \cref{lst:opt-top-level} \cref{line:top-level-cr}}, label={lst:power-law-cr}]
from crikit import CR, TensorType, array
import jax.numpy as jnp

output_type = TensorType.make_symmetric(2, dim)    # Define output and input tensors
input_types = TensorType.make_symmetric(2, dim),   # to be symmetric second-order.
n = array(jnp.array(3))                               # Initialize Glen flow parameter.
def coeff_func(J, n):                                     # Define Wineman-Pipkin
    return jnp.array([0, (J[1] + 1e-12)**((1/n - 1)/2))]) # coefficients for Glen flow.

cr = CR(output_type,input_types, coeff_func, params=(n,)) # Build CR object.
\end{lstlisting}

\begin{lstlisting}[float, caption={Setting up a parametric invariant network CR for \cref{lst:opt-top-level} \cref{line:top-level-cr}.}, label={lst:network-cr}]
import jax
from jax.example_libraries import stax
from jax.example_libraries.stax import Dense, Relu, Tanh
from crikit import cr_function_shape

output_type = TensorType.make_symmetric(2, dim)
input_types = TensorType.make_symmetric(2, dim),
J_len, c_len = cr_function_shape(output_type, input_types)   # Determine number of scalar
                                                             # and form invariants.

init_params_func, predict = stax.serial(Dense(J_len), Relu,  # Build a network with
                                        Dense(5), Tanh,      # one hidden layer.
                                        Dense(c_len))
rng = jax.random.PRNGKey(0)
init_params = init_params_func(rng, (-1, c_len)) # Initialize network parameters.
treedef, flats = tree_flatten(init_params) # Convert parameters into a form
params = [array(x) for x in flats]         # usable by CRIKit.

def coeff_func(invts, *params): # Run network on the scalar invariants.
    return predict(tree_unflatten(treedef, params), invts)
cr = CR(output_type,input_types, coeff_func, params=params)
\end{lstlisting}

\begin{lstlisting}[float, caption={How the CR objects defined in \cref{lst:power-law-cr,lst:network-cr} are evaluated.}, label={lst:cr-evaluation}]
class CR:
  ...
  def __call__(self, inputs, params):
    J = self._scalar_invt_func(*inputs) # Use tabulated functions for these input
    G = self._form_invt_func(*inputs)   #  tensors to compute the invariants.

    vmap_axes = (0,) + (None,) * len(params)          # Compute coefficients and combine
    c = jax.vmap(self._coeff_func, in_axes=vmap_axes)(J, *params) # with form invariants.
    return jax.vmap(jnp.tensordot)(c, G, axes=1) # Calculate Wineman-Pipkin expression.
\end{lstlisting}

\subsection{Code}
The example program in \cref{lst:opt-top-level} shows how the components of CRIKit can be used to define a parameterized CR, simulate an experiment using the CR, optimize its parameters to minimize a loss comparing the simulation to observations, and then adopt the optimized parameters in the CR.
We use FEniCS \cite{alnaes2015fenics,logg2012automated} for expressing and discretizing PDEs, and we use Pyadjoint \cite{mitusch2019dolfin} for composing automatic differentiation operations (reverse-mode for the loss optimization and forward-mode for the nonlinear PDE solver).

CRIKit supports defining a CR by defining the input tensors, output tensor, coefficient function, and CR parameters. 
\Cref{lst:power-law-cr,lst:network-cr} show examples of setting up a CR using CRIKit with the automatic differentiation software JAX by \textcite{jax2018github}. CRIKit currently uses scalar invariants and form invariants tabulated by \textcite{zheng1994theory}. The invariants are looked up based on the input tensors and output tensor.

\Cref{lst:cr-evaluation} shows code for how the invariant CR could be evaluated. The inputs are the state variables at quadrature points. We use JAX's \texttt{vmap} function so that a coefficient function defined for a single quadrature point input can be run on a batch of quadrature points.

\subsection{Stress}

The constitutive relation $r$ relates the strain rate $\dot \epsilon$ to the
deviatoric stress $\tau$, with some parameters $\Theta$.
\begin{align}\label{eq:stress-cr}
\tau &= r(\dot \epsilon; \Theta) \\
\dot \epsilon &= \tfrac{1}{2}(\nabla \vec{u} + \nabla \vec{u}^T)
\end{align}
The standard stress CR for ice flow is Glen's flow law, which is a power law with $n$ usually taken to be 3.
\begin{align} \label{eq:glens-flow-law}
\tau = \mu \sqrt{\dot \epsilon : \dot \epsilon} ^{1/n - 1} \dot \epsilon
\end{align}
\textcite{zheng1994theory} tabulate the scalar invariants $J$ and form invariants $G$ for a wide variety of tensor inputs and outputs. For a tensor function with one symmetric second-order three-dimensional tensor input $A$ and one output of the same description, the scalar invariants can be written as $J(A) = [\tr A, \tr A^2, \tr A^3]$. The form invariants are $G = [I, A, A^2]$, where $I$ is the identity tensor. With these invariants, the Glen flow law in \cref{eq:glens-flow-law} can be expressed in Wineman-Pipkin form with coefficients $c(J(\dot \epsilon)) = [0, \mu J_2^{(1/n - 1)/2}, 0]$. Then $\tau(\dot \epsilon; \mu, n) = \sum_{i=1}^3 c_i(J(\dot \epsilon)) G_i(\dot \epsilon)$.

\subsection{Damage factor}

We use the damage model of \textcite{albrecht2014fracture} to generate our ground-truth observations. In this model, viscosity of the ice is reduced by a damage factor $\phi \in [0, 1]$ that advects with the ice.
\begin{align}\label{eq:damage_stress_cr}
\tau = (1 - \phi) \mu \sqrt{\dot \epsilon : \dot \epsilon} ^{1/n - 1} \dot \epsilon
\end{align}
This is related to the enhancement factor used in other simulations \cite{ma2010enhancement,treverrow_budd_jacka_warner_2012} by $\phi = 1 - E^{1/n}$.
To avoid degenerate equations, they numerically used $\phi_n = (1 - \zeta) \phi$, which bounds $\phi_n$ in $[0, 1-\zeta]$, thereby avoiding the infinite softening problem at $\phi = 1$.

The material time derivative of the damage depends on a fracturing term and a healing term.
\begin{align} \label{eq:damage_rate_cr}
\frac{D\phi}{Dt} = s(\dot \epsilon, \phi) &= s_f + s_h \\
s_f &=
\begin{cases}
\gamma_f \|\dot \epsilon\| (1 - \phi) & \|\dot \epsilon\| > (1-\phi)^{-n} \dot \epsilon_f \\
0 & \text{else}
\end{cases} \label{eq:damage_rate_sf} \\
s_h &=
\begin{cases}
\gamma_h (\|\dot \epsilon\| - \dot \epsilon_h) & \|\dot \epsilon\| \le \dot \epsilon_{h} \text{ and } \phi > 0 \\
0 & \text{else}
\end{cases} \label{eq:damage_rate_sh}
\end{align}
The factors $\gamma_f$ and $\gamma_h$ are scalar constants controlling the rate
of fracturing and healing. The thresholds $\dot \epsilon_f$ and $\dot
\epsilon_h$ control when fracturing and healing begin.
\textcite{albrecht2014fracture} used the local maximum spreading rate as the
norm on $\dot \epsilon$. In this work, we simplify this to the second invariant
$\|\dot \epsilon\| = \sqrt{\dot \epsilon : \dot \epsilon}$, which differs by
at most a factor of $\sqrt{d}$, where $d$ is the dimension.


It is natural to formulate the change in damage as a material derivative that
moves with the ice. For simulation on a fixed mesh, this must be modified to the
Eulerian description below. This is an advective equation with a small amount of
diffusion $\xi$ to make the equation easier to solve.
\begin{align}
\frac{\partial \phi}{\partial t} &= \xi \nabla^2  \phi -
\vec{u} \cdot \nabla  \phi + s
\end{align}
In the equilibrium state, it takes the following form:
\begin{align} \label{eq:damagepde}
\vec{u} \cdot \nabla  \phi - \xi \nabla^2  \phi = s
\end{align}

\begin{figure}
    \begin{subfigure}[t]{0.485\textwidth}
        \centering
        \includegraphics[width=\textwidth]{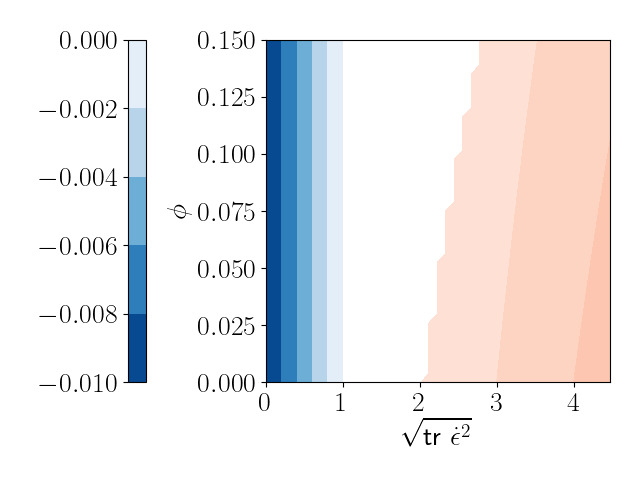}
        \caption{Small strain-rate regime: $\sqrt{\tr \dot \epsilon^2} \in [0,
        \sqrt{20}]$.}
        \label{fig:true_invariant_small}
    \end{subfigure}\hfill
    \begin{subfigure}[t]{0.485\textwidth}
        \centering
        \includegraphics[width=\textwidth]{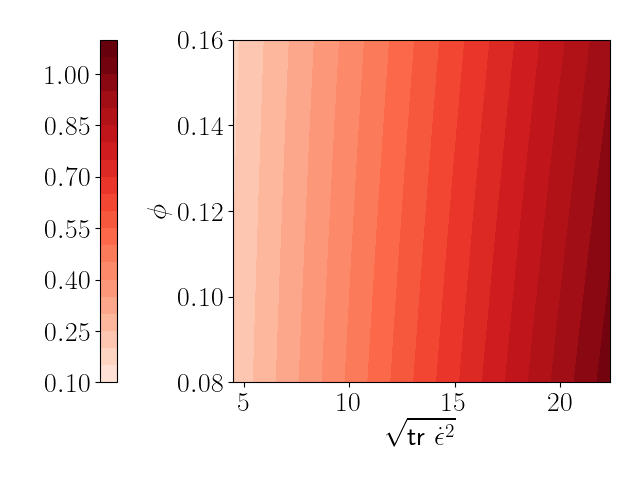}
        \caption{Large strain-rate regime: $\sqrt{\tr \dot \epsilon^2} \in [\sqrt{20},
        \sqrt{450}]$.}\label{fig:true_invariant_large}
    \end{subfigure}
    \hfill
\caption{Ground-truth CR output. When the the strain rate is smaller than the healing threshold $\epsilon_h = 1$, the damage decreases.
There is a discontinuous jump at the fracturing threshold $\epsilon_f = 2$ where the damage rate switches from 0 to $\gamma_f \epsilon_f$.}
\label{fig:ground_truth_cr}
\end{figure}

\section{Numerical experiment}\label{sec:experiment}

\subsection{Boundary conditions}

\begin{figure}
    \centering
    \begin{subfigure}{0.45\textwidth}
    \centering
    \includegraphics[width=\textwidth]{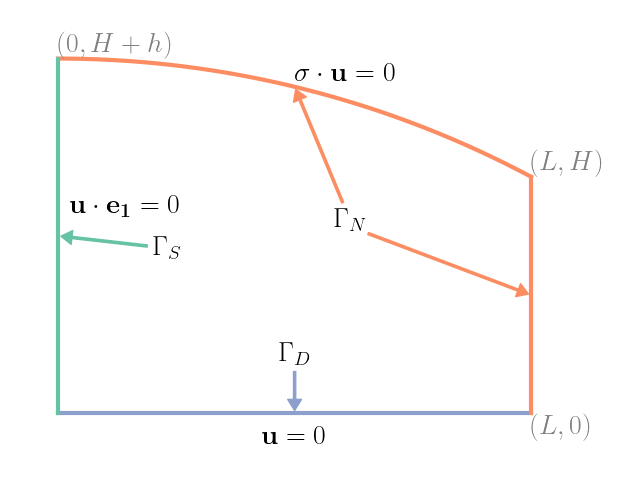}
    \caption{Velocity boundary labels}\label{fig:domain_velocity}
    \end{subfigure}
    \hfill
    \begin{subfigure}{0.45\textwidth}
    \centering
    \includegraphics[width=\textwidth]{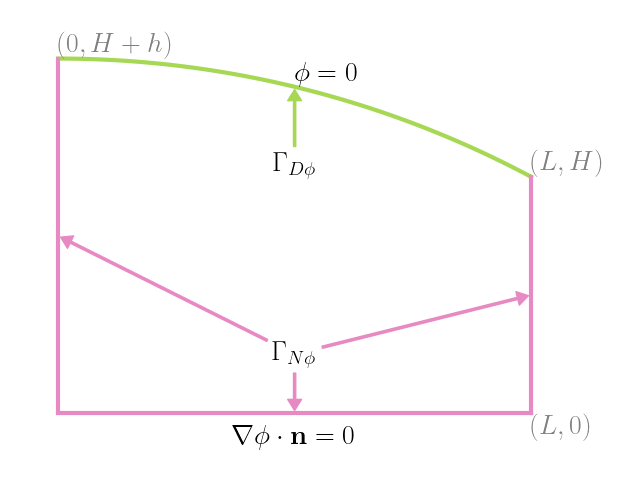}
    \caption{Damage boundary labels}\label{fig:domain_damage}
    \end{subfigure}
    \caption{Domain}\label{fig:domain}
\end{figure}

\begin{figure}
    \centering
    \includegraphics[width=\textwidth]{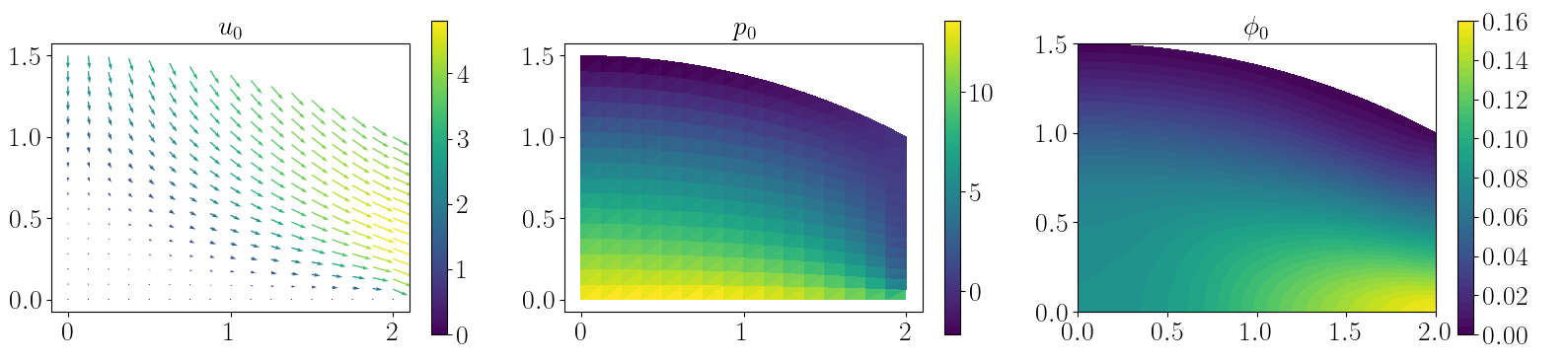}
    \caption{Ground-truth solution}\label{fig:ground_truth_solution}
\end{figure}

We ran a numerical experiment simulating a two-dimensional cross-section of a
grounded ice sheet.
\Cref{fig:domain} shows this cross-section along with
the labeled boundaries and boundary conditions. Since our focus is the
constitutive relation, we used a simplified diagnostic ice sheet simulation. This
included fixed geometry, no-slip boundary at the bed, and stress-free boundaries
at the firn surface and ice front. These boundary conditions are simplified
models that are useful for small tests. The damage is zero at the surface, and
damage flux is zero on the symmetry boundary, the bed, and the calving front.
Zero flux at the calving front is not physically motivated but is useful for a
simple simulation.
\begin{align}
\vec{u} \cdot \vec{e}_1 &= 0 \text{ on $\Gamma_S$} & \text{(symmetry)} \\
\sigma \cdot \vec{n} &= 0 \text{ on $\Gamma_N$} & \text{(stress-free)} \\
\vec{u} &= 0 \text{ on $\Gamma_D$} & \text{(no-slip)} \\
\phi &= 0 \text{ on $\Gamma_{D\phi}$} & \text{(no damage in initial ice)} \\
\nabla \phi \cdot \vec{n} &= 0 \text{ on $\Gamma_{N\phi}$} & \text{(no diffusive flux)}
\end{align}
Here, $\vec{e}_1$ is the horizontal vector
and $\vec{n}$ is the normal vector. \Cref{fig:ground_truth_solution} shows
the solution velocity, pressure, and damage.
The code for running the experiments and generating the plots is publicly available \cite{bruer_grant_2022_6473562}.

\subsection{Discretization}

We express the system using the Galerkin finite-element method with test spaces
$V$, $Q$, and $\Psi$ for velocity, pressure, and damage, respectively. This requires the weak forms of
\cref{eq:force-balance-pde,eq:incompressibility-pde}. Here, $(a,b)$ is defined as the scalar product of $a$ and
$b$ integrated over the domain.
\begin{align}
(\nabla \vec{v}, \tau) -  (\nabla \cdot \vec{v}, p) &= (\vec{v}, \rho \vec{g}) &  \forall \vec{v} \in V \label{eq:weakformv} \\
(q, \nabla \cdot \vec{u}) &= 0 & \forall q \in Q \label{eq:weakformp} \\
(\psi, \vec{u} \cdot \nabla  \phi) + (\nabla \psi, \xi \nabla  \phi)
&= (\psi, s) & \forall \psi \in \Psi \label{eq:continious_phi_weak_form}
\end{align}
\textcite{jouvet2011analysis} showed that the weak formulation of the stationary
Stokes problem with the nonlinear Glen's flow law is well-defined using the following spaces.
\begin{align}
V &= \left\{
\vec{v} \in \left[W^{1,r}(\Omega)\right]^2,\,\, \vec{v} = 0 \text{ on } \Gamma_{D},\,\, \vec{v}\cdot\vec{n} = 0 \text{ on } \Gamma_{N}
\right\}
\\
Q &= L^{r'}(\Omega)
\end{align}
Here, $r = 1 + \frac{1}{n}$, $r' = 1 + n$, and $W^{1,r}(\Omega)$ is the Sobolev space of all scalar function whose value
and derivative are in $L^r$.
We assume $\vec{u} \in
\left[W^{1,r}(\Omega) \cap W^{1,\infty}(\Omega) \right]^2$ and $p \in L^{r'}
\cap L^\infty$.
If damage is fixed and bounded away from 1,
\cref{eq:weakformv,eq:weakformp} are well-posed.  We
say nothing in this work about the well-posedness of the full system over all
possible parameterizations of the damage model $s$, but will discuss how
individual choices of parameters can make the equations difficult to solve in
\cref{sec:results}.

We discretized with piecewise quadratic continuous functions for the velocity and
piecewise constant discontinuous functions for the pressure, $[\mathbb{P}_2]^2 -
\mathbb{P}_0$.
Because the weak form in \cref{eq:continious_phi_weak_form} only requires one derivative, we use a piecwise smooth discretization for $\phi$.
\begin{align}
\Psi &= \left\{
\psi \in H^1,\,\, \psi = 0 \text{ on } \Gamma_{D\phi},\,\, \psi \cdot\vec{n} = 0 \text{ on } \Gamma_{N\phi}
\right\}
\end{align}
We discretize this with $\Psi_h \subset \Psi$ represented by piecewise linear
elements $\mathbb{P}_1$.

We discretized the experiment using the finite-element
package FEniCS \cite{alnaes2015fenics,logg2012automated}.
We use the subscript $h$ to denote the discretized representation.
\begin{align}
(\nabla \vec{v}_h, \tau) -  (\nabla \cdot \vec{v}_h, p_h) &= (\vec{v}_h, \rho \vec{g}_h) &  \forall \vec{v}_h \in V_h \label{eq:weakform_discrete} \\
(q_h, \nabla \cdot \vec{u}_h) &= 0 & \forall q_h \in Q_h \\
(\psi_h, \vec{u}_h \cdot \nabla  \phi_h) + (\nabla \psi_h, \xi \nabla  \phi_h + \delta_h R \vec{u}_h)
&= (\psi_h, s) & \forall \psi_h \in \Psi_h \label{eq:phi-advect-diffuse}
\end{align}
The term $(\delta \vec{u}_h, R \nabla \psi_h)$ is a stabilizing streamline-upwind
Petrov-Galerkin (SUPG) term, where $\delta_h$ is an $O(h)$ function of element
size, and $R = \vec{u} \cdot \nabla \phi - \xi \nabla^2 \phi - s$ is the
strong-form residual.

\paragraph{Damage factor bound}
The material derivative for the damage evolution ensure $\phi \in [0, 1]$, but
the system becomes singular when $\phi = 1$, which corresponds to zero viscosity
in \cref{eq:damage_stress_cr} (the stress CR). Following \textcite{albrecht2014fracture},
we replace $\phi$ in \cref{eq:damage_stress_cr} with
$\widetilde \phi = 0.999 \phi$ to ensure the damage is bound below zero.

\paragraph{Neural network well-posedness}
When inferring the constitutive relation with a neural network instead of Glen's
flow law, we assume the problem is well-posed. When the system is ill-posed
(e.g., we encounter a singular Jacobian), we define the loss to be $\infty$.
When the problem is ill-conditioned (e.g., large damages causes large velocities
which make the advective system difficult to solve), the relative residual after
running a solver may be larger than the desired residual tolerance.
In either case, the solution returned by the
solver will not be close to the ground-truth, and the loss will be large. Thus,
when the optimizer starts to enter a bad area in parameter space, a line search
will cause it to back out of that area.

The decision on whether to set the loss to $\infty$ or to use the solution
returned by the solver can be made based on the relative residual. An ill-posed
system tends to diverge and fail the Jacobian solve, whereas an ill-conditioned
system converges to a residual that doesn't satisfy the specified tolerance. In
practice, it's easier to not differentiate between these cases and instead simply
set the loss to an extremely large value if the system fails to solve to the
desired tolerance.

Note that this is not ideal. Neural networks are good function approximators,
and when initialized with random values, they can approximate a lot of different
functions. By backing away from bad solutions, we limit the parameters to ones near
the initialization, so it may be possible that there's an ill-posed system blocking
the parameters from the best approximation. We do not investigate this effect here
and leave this concern to a future work.

This brings a caveat to the transfer learning idea. If we do transfer learning
based on existing models, then we are biasing the training procedure to look
like the initial models. Since there has to be a path of solvable systems
between the parameter initialization and the final trained parameters, the final
results are more similar to evolutions of the initial models to fit the data, instead of
the best possible solution out of all solutions.

We leave the proof for the necessary constraints on the neural network for
solvability and convergence to a future work.

\subsection{Loss function}

\subsubsection{Invariant loss}
A standard loss function for regression is an $l_2$ measure of the error. If we
have a set of measurements $M = \{J^1, J^2, \ldots \}$ of the input invariants
$J(\epsilon, \phi)$ and the corresponding output of the ground-truth function
$s_0$, then we can train our network function $s$ on the data using the direct invariant loss
function $L_d$.
\begin{equation}\label{eq:invariant_loss}
L_d(s, s_0) = \sum_i
(s(J^i) - s_0(J^i))^2
\end{equation}
With real-world ice sheet data, we won't have access to these direct input-output pairs,
but this loss is still useful for measuring network performance in our simulations.

\begin{figure}
    \begin{subfigure}[t]{0.485\textwidth}
        \centering
        \includegraphics[width=\textwidth]{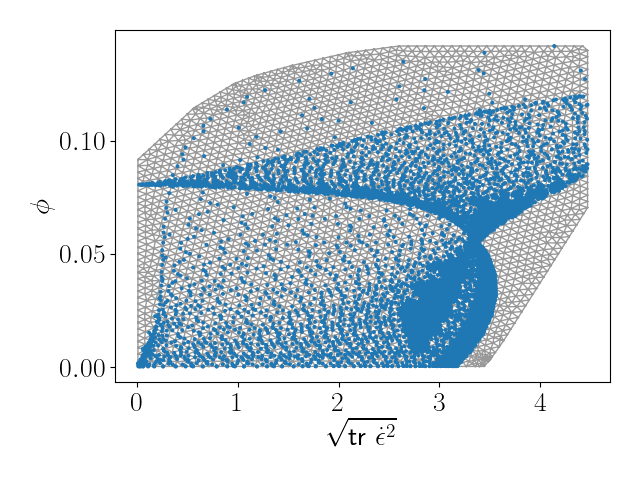}
        \caption{Small regime: $\sqrt{\tr \dot \epsilon^2} \in [0, \sqrt{20}]$.}
    \end{subfigure}\hfill
    \begin{subfigure}[t]{0.485\textwidth}
        \centering
        \includegraphics[width=\textwidth]{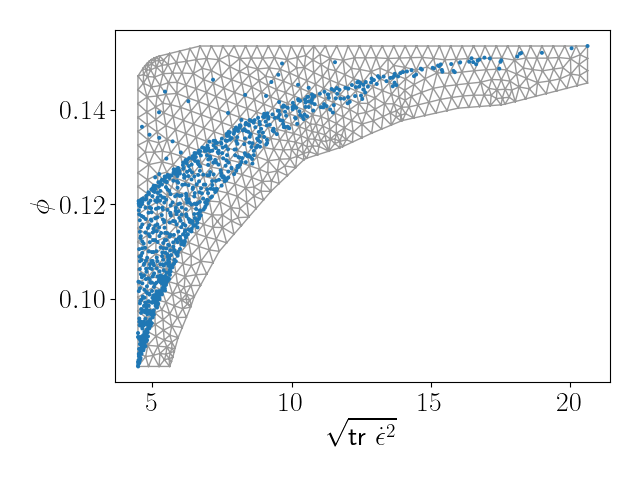}
        \caption{Large regime: $\sqrt{\tr \dot \epsilon^2} \in [\sqrt{20}, \sqrt{450}]$.}
    \end{subfigure}
    \hfill
\caption{The blue points show the location of the invariants in the experiment. The mesh encompasses these points and can be used as the domain for the direct invariant loss in \cref{eq:invariant_loss}.}
\label{fig:invariant_space_mesh}
\end{figure}

The data points $\{J^i\}$ can be chosen from some invariant-space domain of
interest, e.g., a mesh over the domains shown in
\cref{fig:invariant_space_mesh}. Or they could instead be computed
at points in the physical domain as $J^i = (J_1(\vec{x}^i), J_2(\vec{x}^i))$,
which drastically changes the distribution of training values.
\Cref{fig:invariant_space_mesh} shows the invariant-space domain
with points plotted at values that show up in the experiment. The concentration
of the points is much higher in the small regime and very sparse at the highest
damage values.

The choice of invariant data presents tradeoffs.
On the one hand, a network trained on invariants uniformly sampled from a
region of invariant-space would perform better over a wider range of experiments.
On the other hand, a network trained on the invariant values computed at points
in the physical space mesh would better learn the distribution of real data and
would perform better at areas in invariant-space with high concentrations of data
that shows up in the experiment.

Ideally, we would have a network that performs well with small perturbations of
the experiment. For our invariant loss, we focus on generalizability of the
resulting network and therefore use an integral over invariant-space domain. We
choose the mesh to extend slightly outside the range of data that shows up in
the experiment.

\subsubsection{Experimental loss}
Despite the preceding discussion, the outputs and inputs to the CR--the stress, strain, and damage--cannot be
directly measured easily, if at all. Instead, the measurable quantities for this
system are the velocity $\vec{u}$ and the pressure $\vec{p}$. The loss function for training
should only depend on these observable solutions of the governing PDE.

If we have measurements over the full interior of the domain, we can do a simple
$L_2$ norm of the error to use for the loss function.
\begin{equation}\label{eq:experimental_loss_interior}
L_i(\vec{w}, \vec{w}_0) = \int_\Omega{\|\vec{u} - \vec{u}_0\|_2^2 + \|p - p_0\|_2^2} \,\,dx
\end{equation}
where $\vec{w} = (\vec{u}, p)$ is the simulated observations using the candidate
CR $s$, and $\vec{w}_0 = (\vec{u}_0, p_0)$ is the measured velocity and pressure
fields produced by the target unknown CR $s_0$.

Obtaining measurements over the full interior in a real ice sheet or glacier is infeasible.
Generally measurements of ice sheets are restricted to surface measurements,
possibly augmented with borehole measurements.
The surface's contribution to the
loss can be represented as a surface integral over the top of the domain
$\Gamma_T$, while a borehole can be represented as a line integral in the
interior of the domain.  To implement borehole measurements in a mesh-agnostic
way, we instead approximate them using a volume integral over a
narrow domain $\Omega_B$. These domains (and other useful partitions) are
labeled in \cref{fig:ground_truth_cr_results}.
\begin{align}
L_s(\vec{w}, \vec{w}_0) &= \int_{\Gamma_T}\|\vec{u} - \vec{u}_0\|_2^2 + \|p - p_0\|_2^2 \,\,ds
\\ L_b(\vec{w}, \vec{w}_0) &= \int_{\Omega_B} \gamma_u\|\vec{u} - \vec{u}_0\|_2^2 + {\gamma_p}\|p - p_0\|_2^2 \,\, dx
\end{align}

We investigate the performance of our networks by training on the full interior
loss $L_i$, the surface loss $L_s$, or the combined surface borehole loss $L_s +
L_b$.
Scaling factors are necessary when summing error contributions with different
units, and we choose scaling factors $\gamma_u$ and $\gamma_p$ to make the
contributions from the borehole velocity and pressure terms approximately equal.
For simplicity, we include the surface pressure in the loss although to $O(h)$ it should be approximately 0.

\subsection{Noise}

In our simulation, we know the correct CR and can simulate
$\vec{w}_0$ precisely, but real-world measurements are subject to noise. We
reason that zero velocities should be measured essentially exactly, whereas
larger velocities will have a larger noise contribution. This reasoning does not
apply to pressure, since pressure is unique only up to a constant. Therefore, to
simulate noise, we draw from a Gaussian distribution with standard deviation
equal to a proportion of the range of each value, using multiplicative noise on
the velocity and additive noise on the pressure.

\begin{align}
\widetilde{\vec{u}}_0(\vec{x}) &= \vec{u}_0(\vec{x}) + \delta_u(\vec{x}) \vec{u}_0(\vec{x}) \\
\widetilde{p}_0(\vec{x}) &= p_0(\vec{x}) + \delta_p(\vec{x}) (\max_{\vec{x} \in \Omega} p_0(\vec{x}) - \min_{\vec{x} \in \Omega} p_0(\vec{x}))
\end{align}
where $\delta_u(\vec{x}) \sim \mathcal{N}(0, \delta)$ and $\delta_p(\vec{x}) \sim
\mathcal{N}(0, \delta)$ for noise proportion $\delta$. We test with 0\%, 1\%, and 5\% noise.

\subsection{Data collection}

\begin{figure}
    \centering
    \includegraphics[width=0.8\textwidth]{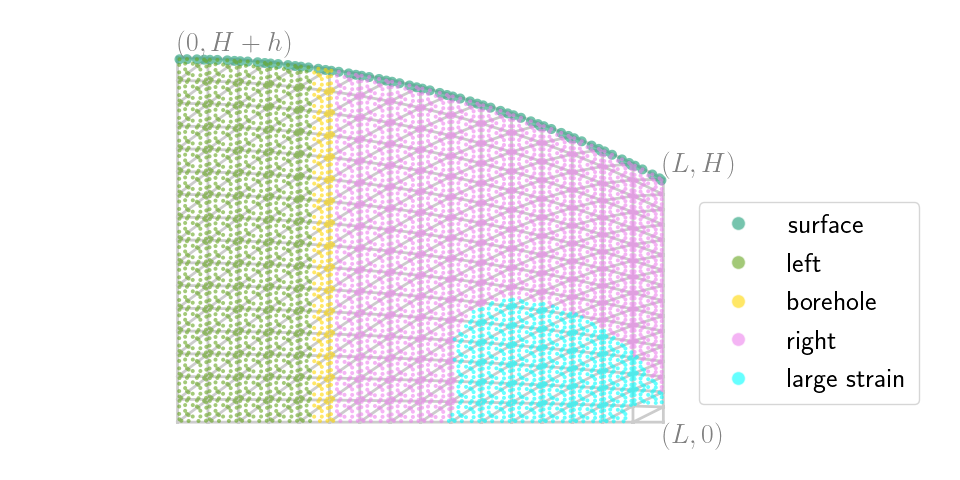}
    \caption{The points in the CR plots are colored according to where the
    quadrature points fall in the physical domain. The bottom right of the
    physical domain is a large strain rate regime, which is displayed
    separately in the following plots since it exists on a wildly different
    scale in invariants space. Furthermore, the bottom right corner of the
    domain is removed from consideration because the strain rate is
    incredibly large there due to numerical artifacts of the two different
    boundary conditions at the corner.}\label{fig:domain_mesh_partitions}
\end{figure}

\begin{figure}
    \begin{subfigure}[t]{0.485\textwidth}
        \centering
        \includegraphics[width=\textwidth]{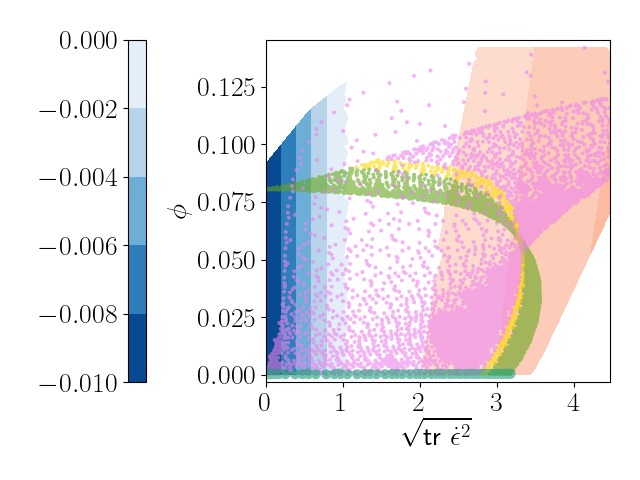}
        \caption{Small strain-rate regime: $\sqrt{\tr \dot \epsilon^2} \in [0,
        \sqrt{20}]$. Note that the points on the surface of the physical domain are
        shown on the bottom edge of this plot, because the damage is 0 along the
        surface.}
        \label{fig:true_invariant_small_partitions}
    \end{subfigure}\hfill
    \begin{subfigure}[t]{0.485\textwidth}
        \centering
        \includegraphics[width=\textwidth]{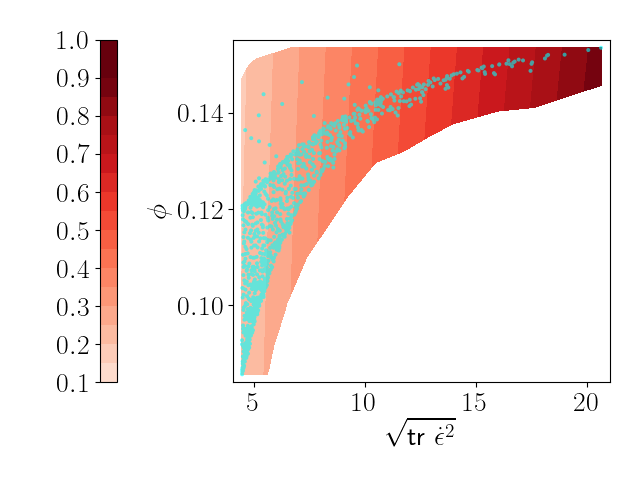}
        \caption{Large strain-rate large regime: $\sqrt{\tr \dot \epsilon^2} \in [\sqrt{20},
        \sqrt{450}]$.}\label{fig:true_invariant_large_partitions}
    \end{subfigure}
    \hfill
\caption{Ground-truth CR output $s_0$ with invariants that show up in the experiment marked.}
\label{fig:ground_truth_cr_results}
\end{figure}

Our goal is to examine how well a randomly initialized network can converge to a
good solution and to examine the effect of noisy/limited observations on this
training.
%
%
Since we use the Wineman-Pipkin representation, the direct inputs to the
CR are the scalar invariants, and the output is the damage rate.
\Cref{fig:ground_truth_cr_results} shows the output of the ground-truth CR
from \cref{eq:damage_rate_cr} in this CR domain, both in a small strain-rate
regime and a large strain-rate regime. The plotted points show the conditions
that occur in the ground-truth experiment. Specifically, these points show the
values of the scalar invariants at the quadrature points in the physical domain
mesh used for the FEM integration; their colors correspond to \cref{fig:domain_mesh_partitions}.


The inputs to the network are scaled to have zero mean and unit variance, based
on the mean and variance of the invariants that show up in the ground-truth
solution. The network weights are initialized using a standard normal
distribution, and the biases are initialized to zero. \Cref{tbl:hyperparams}
shows the hyperparameters that we tested, which fall into the following categories:

\paragraph{Network structure:} The network has two inputs and one output. We
test using one layer, two layers, and three layers, with each layer being
fully-connected. For each network depth, we use either two units or four units
for all the hidden layers.

\paragraph{Activation function:} Hyperbolic tangent (tanh) is a classic choice
of activation function, while ReLU is a activation type that avoids some of
tanh's vanishing gradient issues. We also test softplus activation, which is a
smooth version of ReLU, and since it has a nonzero gradient everywhere, it can
better recover from poor initializations.

\paragraph{Optimizer:} We tested a standard BFGS optimizer along with a trust
region version of BFGS (both implemented in SciPy). The idea was that the trust
region may help keep the network from getting to unsolvable locations, thereby
making the optimization run more smoothly.

\begin{table}
\centering
\begin{tabular}{ccccccccc}
&Network structure &&&&
Activation &&&
Optimizer algorithm \\
\cline{1-3}
\cline{6-6}
\cline{9-9}
2 \textrightarrow & 2 & \textrightarrow 1 &&& tanh &&& BFGS \\
2 \textrightarrow & 2 \textrightarrow 2 & \textrightarrow 1 &&& ReLU &&& BFGS with trust region \\
2 \textrightarrow & 2 \textrightarrow 2 \textrightarrow 2 & \textrightarrow 1 &&& softplus \\
2 \textrightarrow & 4 & \textrightarrow 1 \\
2 \textrightarrow & 4 \textrightarrow 4 & \textrightarrow 1 \\
2 \textrightarrow & 4 \textrightarrow 4 \textrightarrow 4 & \textrightarrow 1
\end{tabular}
\caption{Network and optimizer hyperparameters that we tested.}\label{tbl:hyperparams}
\end{table}

\subsection{Results}\label{sec:results}

The effectiveness of our method for learning network parameters is affected in sometimes subtle ways by the choices that go into the loss and the optimization procedure. 


\subsubsection{Network hyperparameters}

\paragraph{Network structure}

\begin{figure}
    \begin{subfigure}[t]{0.9\textwidth}
        \centering
        \includegraphics[width=\textwidth]{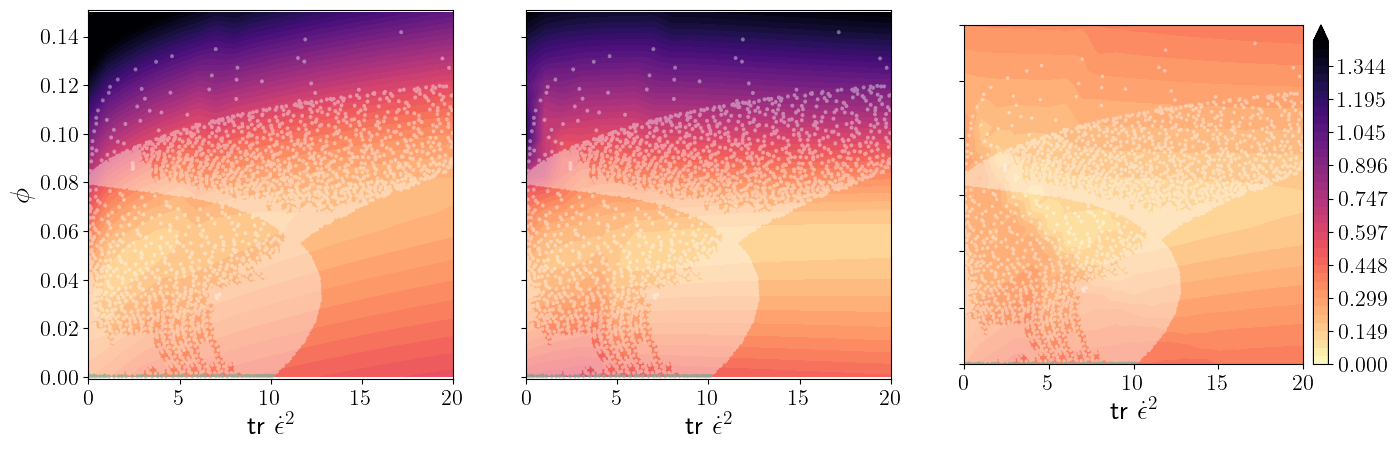}
        \caption{From left to right: (2,), (2, 2), (2,2,2)}
    \end{subfigure}
    \begin{subfigure}[t]{0.9\textwidth}
        \centering
        \includegraphics[width=\textwidth]{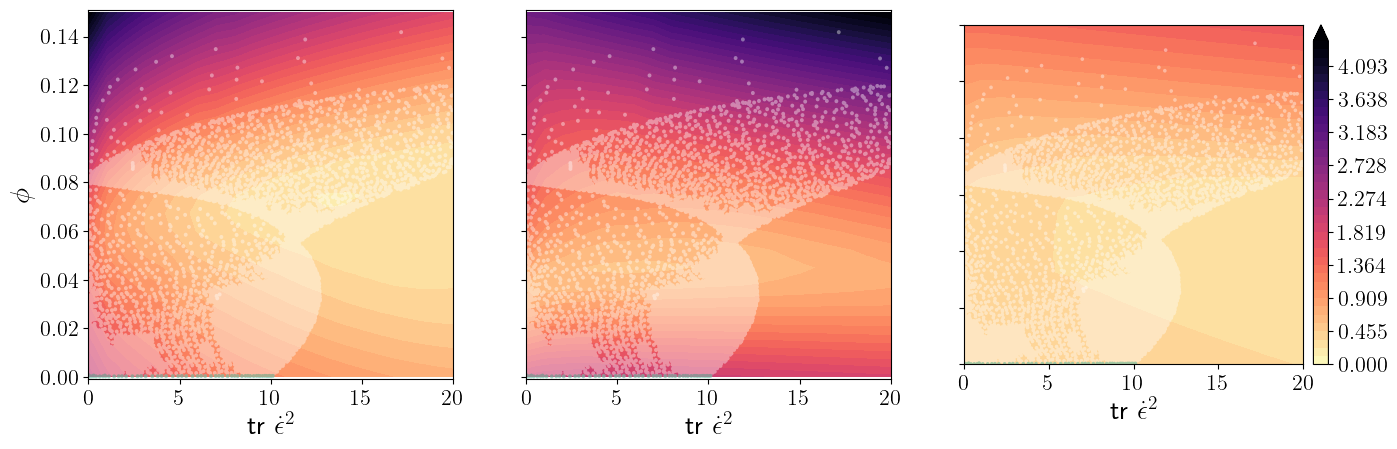}
        \caption{From left to right: (4,), (4, 4), (4,4,4)}
    \end{subfigure}
\caption{Network size results: example RMSE comparison of CRs in CR domain,
showing the trend that the deeper network did better for layer size 2 and that
layer size 4 had more difficulty with (4, 4) than the (4,) and (4, 4, 4). These networks
used softplus activation and were trained on surface observations with BFGS. \todo{this figure is good since it shows in the absence of noise, we get some positive effect of increasing network size. We can probably do these on the same color scale and make them show what we want.}}
\label{fig:ex_net_cr_domain}
\end{figure}

\Cref{fig:ex_net_cr_domain} plots the RMSE on the invariant domain for a representative network activation and observation type. With the right choice of weights, a network that is wider and deeper can express the same function as a smaller network, so we expect the deeper and wider networks to perform as well or better than the smaller networks. Indeed, the deepest networks we tested were able to achieve a lower invariant loss over a larger region of the invariant domain. However, the wider networks often performed worse than the less wide networks of the same depth. This indicates the larger networks are not trained as well as the smaller networks, likely due to a more complex loss landscape.

\paragraph{Activation function:}

We found very little difference between the loss value of the ReLU networks
compared to the softplus networks. The notable difference is a small number of
the ReLU networks converged to output a constant across the entire domain. This
is less likely to happen with softplus networks because softplus activation has
a nonzero derivative everywhere. For that reason, we exclude the ReLU activation
from the results below. Regarding the softplus and tanh networks, we don't
observe much difference between them, so we present them together in what
follows.

\paragraph{Optimizer:}
We found very little difference between the two optimization methods, both in
regards to the experimental loss and the invariant loss. Standard BFGS tended to
perform slightly better, so for the sake of space, we show just the standard
BFGS results.

\subsubsection{Initial solvability}

With our choice of random initial weights drawn from a standard normal distribution, we find that smaller networks are more likely to yield solvable PDE systems than larger networks, and networks
using hyperbolic tangent activation are slightly more likely to be initially solvable than softplus or ReLU.
This effect of network size and activation function on the solvability
can be explained by examining the magnitude of the damage rate, and the explanation for this issue led us to a strategy for obtaining a network with valid initial weights.

The easiest way for the PDE to become difficult to solve is in the damage
advection equation. The system will not converge if the advection term is too
large relative to the stabilizing diffusive term.
For a fixed strain rate, a higher damage rate leads to
higher damage which leads to lower stress. This effect
causes a larger velocity, which tends to make the advective system more ill-conditioned.
With this explanation, we see that the hyperbolic tangent activation function can better avoid ill-conditioned systems because it keeps the activations between -1
and +1 in every hidden layer. This caps the damage rate at a level controlled by
the weights of the final linear layer. In contrast, the softplus and ReLU
activations can have unbounded growth in the damage rate, which worsens as the number of layers is increased.

To fix this issue, we consider that when the network weights are all zero, the damage rate is zero.
The resulting PDE is a standard Glen flow system, which doesn't have the issue of damage advection being ill-conditioned. If we consider zero weights as an initial known solvable state, we can think of choosing initial random weights as picking a random search direction in an optimization routine. A full step in the random direction yields an infinite loss, so we decrease the size of the step until the resulting PDE system is solvable, analogous to a backtracking line search in an optimization routine.

\subsubsection{Solvability during optimization}

We can use the loss gradient to ascertain how close we are to a local minimum in the loss because the gradient should be approximately 0 when optimization is complete.
However, the only runs that achieved a minuscule loss gradient ($\approx 10^{-14}$) were poorly initialized ReLU networks. These networks gave a constant output independent of the input because they were stuck in the flat region of the ReLU curve.

All of the other networks had a final loss gradient with a magnitude on the same
order of the loss. From sensitivity analysis of some of these trained networks,
we find that a small step in the direction of the gradient does reduce the loss,
so standard gradient descent could still achieve a smaller loss. The issue is
that the step suggested by BFGS reaches an unsolvable state with a large loss,
which causes the cubic line search to make the step tiny. This makes no change
to the loss and causes the optimization to terminate.

\subsubsection{Observations and noise}

\begin{figure}
\centering
\includegraphics[width=\textwidth]{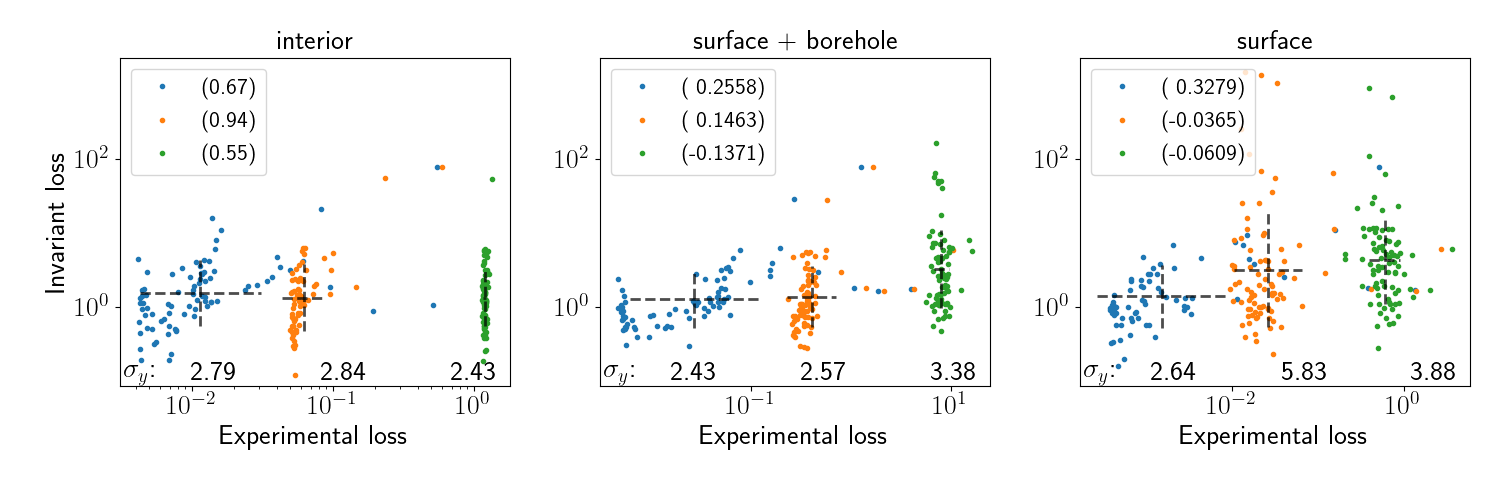}
\caption{Plots showing the correlation between experimental loss and invariant loss for the trained networks. Each color shows a different noise level.}
\label{fig:correlation}
\end{figure}

\Cref{fig:correlation} shows the correlation between the experimental loss and the invariant loss for each of the three observation types.
A wide variety of invariant losses are obtained for networks with approximately the same experimental loss.
From left to right, the three plots show decreasing information in the loss. The main difference between the results for the different observation types is that taking few observations leads to a much larger spread in the experimental loss for the noisy runs.

For the noiseless data (shown in blue), each has approximately the same mean of invariant loss, which is shown as the horizontal dashed line over the blue point group in each plot. This is likely due to the advective nature of the damage field, which causes non-local effects. Even if measurements are taken only on the surface, with noiseless measurements, the interior damage CR can be estimated well.

The noise level affects the mean invariant loss and standard deviation compared across measurement types. For 1\% noise, interior and surface+borehole achieved the same invariant loss, whereas the surface-only version had a larger mean invariant with a much larger multiplicative standard deviation. For 5\% noise, the effect on surface+borehole compared to interior is more pronounced, but surface+borehole is still noticeably better than surface-only.

\begin{figure}
\centering
\includegraphics[width=\textwidth]{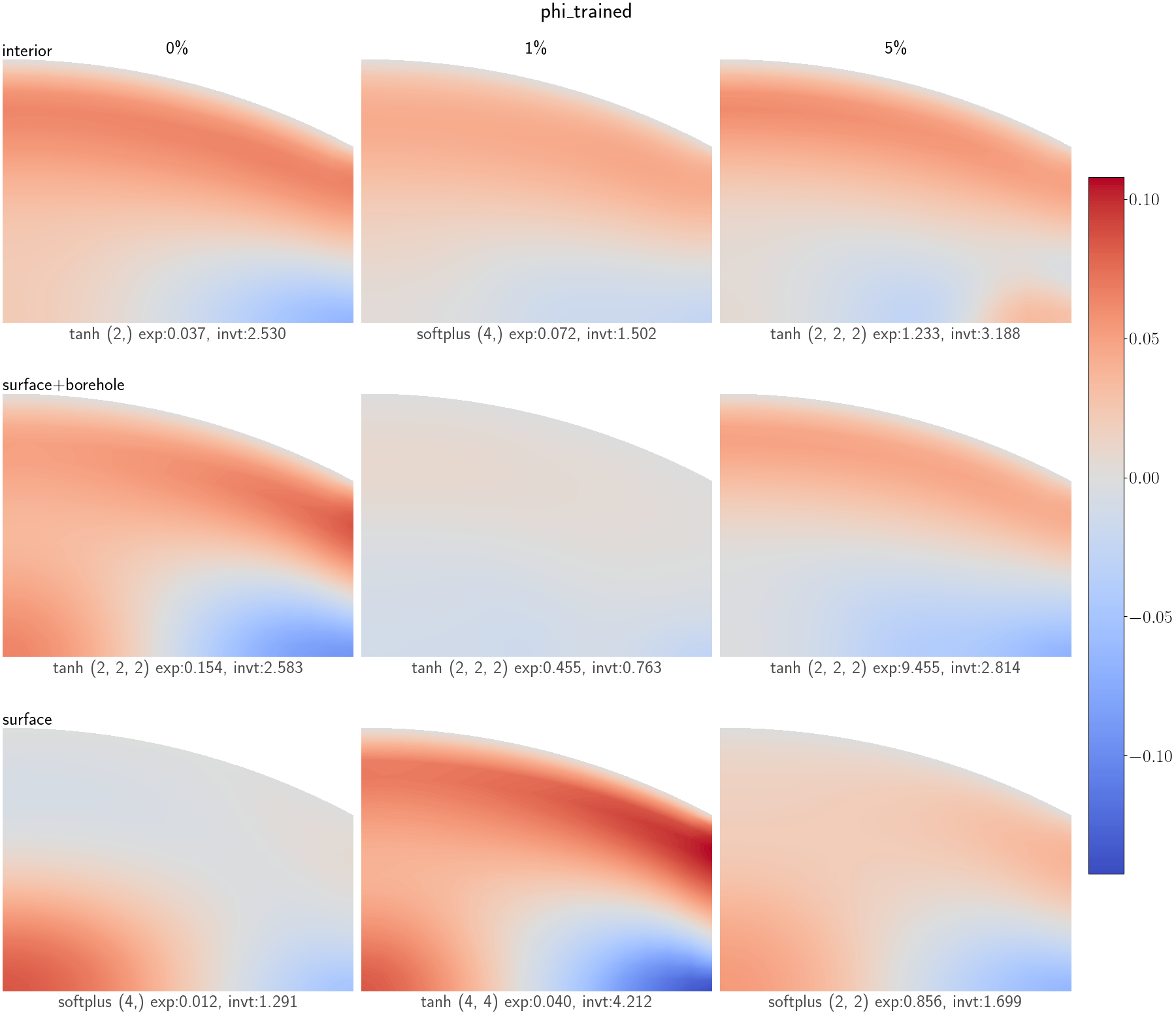}
\caption{For each noise level and each observer type, this figure shows the error
in $\phi$ in the physical domain for the network with the lowest experimental loss. \todo{Maybe make the color bar be evenly split between negative and positive values. Maybe use coarser color granularity so that it is easier to compare colors and see contours.} }
\label{fig:delta_phi}
\end{figure}

To see this pronounced difference in the different observers, we examine each observer and noise level pair and plot the difference in the true damage field and the predicted damage field by the network that achieved the lowest loss. This is shown in \Cref{fig:delta_phi}.


\begin{figure}
\centering
\includegraphics[width=\textwidth]{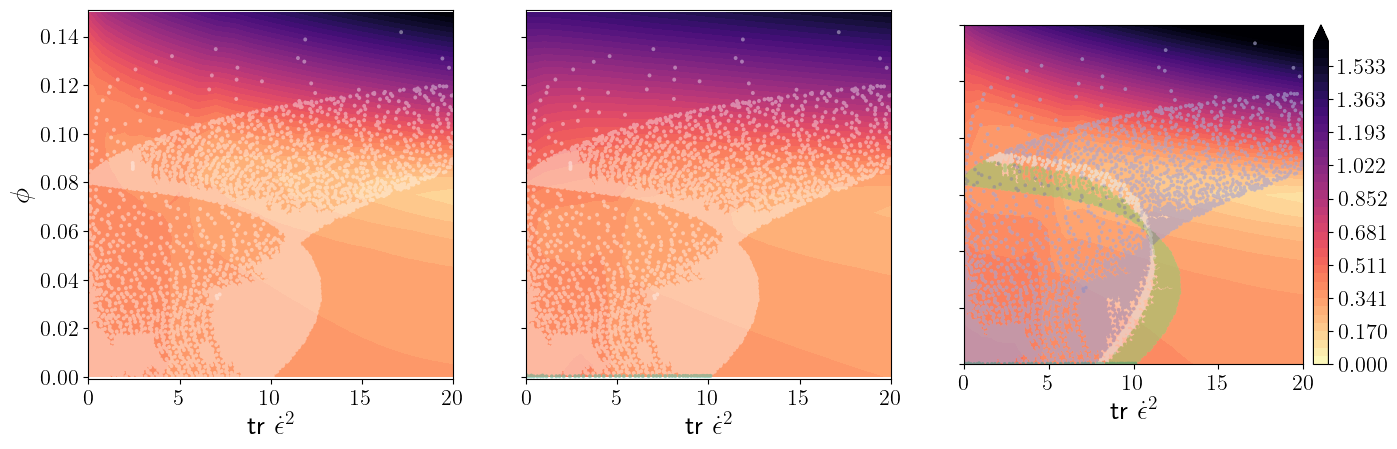}
\caption{Example observer results in CR domain. From left to right: interior, surface, surface+borehole.
There are only slight differences between the three. These networks were ``Softplus  (4, 4, 4)" and were trained with
BFGS. \todo{replace this with just the one's error}}
\label{fig:ex_obs_cr_domain}
\end{figure}

\begin{figure}
\centering
\includegraphics[width=\textwidth]{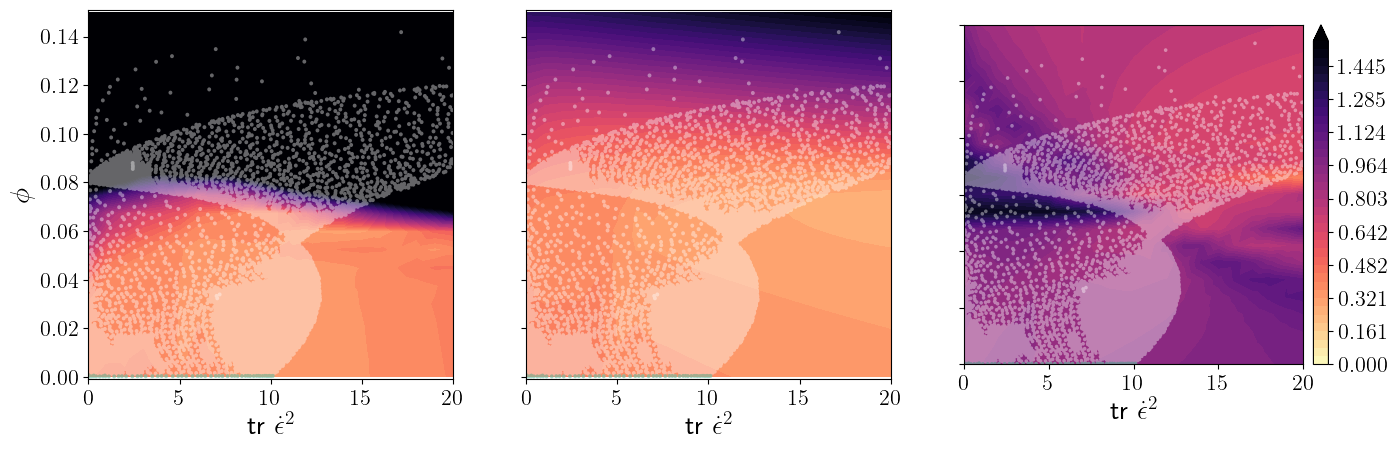}
\caption{Activation RMSE results in CR domain. From left to right, these are ReLU, softplus, and tanh.
The trend is that the ReLU does very poorly. Softplus tends to have the smallest
errors. These networks were ``(4,4,4)" and were trained on surface observations
with BFGS. \todo{demonstrates the non-differentiability of relu leads to bad rmse.}}
\label{fig:ex_act_cr_domain}
\end{figure}

\section{Summary and Discussion}

Our experiences in training neural network CRs that reproduce the Albrecht-Levermann damage model from quasi-realistic observations are well summarized by the correlations between the experimental loss and invariant loss illustrated in \cref{fig:correlation}.  Several issues that affect the viability of this method can be seen therein.  In this section we present these issues together with some potential solutions.

There is significant spread in the experimental loss (the horizontal spread in \cref{fig:correlation}) for optimizations from different randomly initialized parameters, even
for data generated without noise and with observations throughout the interior of the ice sheet.  This illustrates that the loss landscape for the network parameters is both highly non-convex, with multiple local minima, and also that the random initialization of parameters results in many network CRs for which the experiment equations \cref{eq:incompressibility-pde,eq:force-balance-pde} are not solvable.

Ideally we should be able to initialize network parameters in such a way that
"realistic" parameters (those for which the equations are solvable) are easily
generated.  One way to do this would be to project existing, hand-derived CRs into
the parameter space using, for example, the invariant loss to define the projection.
If we were to do this deterministically, however, we might worry that the trained CRs
might converge only to local minima in the vicinity of those projections.  One way to
allow hand-derived CRs to influence the initialization of network CRs non-deterministically would be to define prior distributions over their parameters and then project them into the
neural network parameterization.  So, for example, in the Albrecht-Levermann damage model, one could define physically realistic distributions of the parameters $\gamma_{f}$ and $\gamma_{h}$ and create an initial guess by randomly sampling the two and finding the closest network CR according to the invariant loss.

Even when networks train consistently to an optimal value for the experimental loss (such as in the high-noise, interior observations data in \cref{fig:correlation} (left, green)),
there remains significant variation in the value of the invariant loss for the optimized network CRs.  This is characteristic behavior for an ill-posed inverse problem, when
the parameterization is essentially under-determined.

Standard approaches such as Tikhonov or $\ell_1$ regularization of network parameters
can make the inference problem well-posed, but the data-theoretic complexity that they
encode does not map directly onto the complexity of the target invariant response of the network CR.  One way of encoding this would be an $H^1$-Tikhonov or total variation regularizer in invariant space, essentially penalizing how changeable the CR for similar inputs.  It is also the case that additional known features of the damage-model CR have not been imposed on the network structure of network CRs tested in our experiments, and that incorporating those features into the network design would lead to more consistent training. For instance, in the Albrecht-Levermann model, and in most \todo{check this} of the other models listed in \cref{sec:model_survey}, the rate of damage is non-decreasing in some measure of strain rate (as visible in \cref{fig:ground_truth_cr_results}).  Network architectures exists to encode monotonicity of the response to subsets of inputs \cite{lang_2005}, and it could be that restricting the network CRs in this way would be beneficial.

\printbibliography

\appendix

\section{Wineman-Pipkin examples}
\label{app:wineman-pipkin}

In Wineman-Pipkin form, each constitutive relation is represented in invariant
form by computing form invariants, scalar invariants, and form coefficients that
are functions of the scalar invariants.

Let $P$ be the tuple of input tensors, $G(P)$ be the tuple of form invariants,
$J(P)$ be the tuple of scalar invariants, and $c(J)$ be the form coefficients
for the stress. Then, for any constitutive relation function $f(P)$, we can write it in the following form:
\begin{align}
f(P) = \sum_i c_i(J(P)) \,\, G_i(P)
\end{align}
The choice of form invariants and scalar invariants is not unique. For many different types of inputs, \textcite{zheng1994theory} has recorded a set of scalar invariants as well as a set of form invariants that have the desired structure for the output. Here, the output is the stress, which is a symmetric second-order tensor.

\paragraph{Glen flow}
For example, for Glen flow, we have one input $P = \dot \epsilon$. The input and output are both symmetric second-order tensors. \citeauthor{zheng1994theory} records the form
invariants in 2D to be $G(P) = [I, \dot \epsilon ]$ with scalar invariants as $J(P) =
[\tr \dot \epsilon, \sqrt{\tr \dot \epsilon^2}]$. Then the form coefficients
are $c(J) = [0,\mu J_2^{1/n - 1}]$, where $\mu$ and $n$ are model parameters,
\begin{align}
\tau(P) &= \sum_i c_i(J(P))\, G_i(P) = \mu \sqrt{\dot \epsilon : \dot \epsilon} ^{1/n - 1} \dot \epsilon.
\end{align}

\paragraph{ESTAR flow relation}
ESTAR described by \textcite{graham2018implementing} parameterizes the enhancement factor by expressing it as a variation between two constant enhancement factors $E_C$ (compression) and $E_S$ (shear).
These are the enhancement factors for isotropic ice either under compression or simple shear.
The enhancement factor at a given location in the ice varies based on the proportion of shear stress $\lambda_S$.
With the decomposition $\tau = \tau' + \tau^{\perp}$,  $\lambda_S$ is expressed as $\lambda_S = \|\tau'\|/\|\tau\|$, where $\tau'$ is the shear stress acting on the locally non-rotating shear plane,.
The magnitude of shear stress $\tau'$ is calculated using the magnitude of shear strain rate $\dot \epsilon'$ on the locally non-rotating shear plane.
The vorticity $\omega = \nabla \times \vec{u}$ can be decomposed into a rigid-body rotation component $\omega_R$ and a deformation component $\omega_D$ with unit vector $\hat{\omega}_D$.
The normal $\vec{n}$ to the local shear plane is computed using $\hat{\omega_D}$.
The deformational vorticity is approximated as $\overline \omega_D$; this does not include the components parallel to the flow $\vec{u}$, which are assumed to be small.
Given the inputs $P = [\dot \epsilon, \nabla \times \vec{u}, \vec{u}]$, the stress can be computed using

\begin{align}
\vec{\omega}_1 & \gets \nabla \times \vec{u} - \frac{2\vec{u} \times ((\vec{u} \cdot \nabla) \vec{u})}{\|\vec{u}\|^2} \label{eq:omega_1},
\\
\overline \omega_D &\gets \vec{\omega}_1 - (\vec{u} \cdot \vec{\omega}_1)\frac{\vec{u}}{\|\vec{u}\|^2} \label{eq:omega_D},
\\
\hat{\omega}_D &\gets \frac{\overline \omega_D}{\|\overline \omega_D\|} \label{eq:omega_D_hat},
\\
\vec{n} &\gets \frac{\vec{u} \times \overline {\vec{\omega}}_D}{\|\vec{u} \times \overline {\vec{\omega}}_D\|} \label{eq:shearplane},
\\
\dot \epsilon' &\gets \dot \epsilon \vec{n} - (\vec{n}^T \dot \epsilon \vec{n}) \vec{n} -
(\hat{\vec{\omega}}_D^T \dot \epsilon \vec{n}) \hat{\vec{\omega}}_D \label{eq:shearstrain},
\\
\lambda_S &\gets \frac{\|\tau'\|}{\tau_e} \gets \frac{\|\dot \epsilon'\|}{\sqrt{\dot \epsilon : \dot \epsilon}} \label{eq:lambda_s},
\\
E &\gets E_C + (E_S - E_C) \lambda_S^2 \label{eq:elambda_s},
\\
\tau &\gets E(\lambda_S)^{1/n} \sqrt{\dot \epsilon : \dot \epsilon} ^{1/n - 1} \dot \epsilon \label{eq:estar_tau}.
\end{align}
Each of \cref{eq:elambda_s,eq:lambda_s,eq:shearstrain,eq:shearplane,eq:omega_1,eq:omega_D,eq:omega_D_hat} can be expressed in Wineman-Pipkin form, which implies \cref{eq:estar_tau} can be expressed in Wineman-Pipkin form.

\paragraph{Damage}
This model is described by \textcite{albrecht2014fracture}. They include a scalar state variable $\phi$ representing the ice damage, which advects with the ice. The advection equation requires a closure $s$ representing the damage rate. The inputs to the closures for stress and damage rate are $P = [\dot \epsilon, \phi ]$, which have scalar invariants $J(P) = [\tr \dot \epsilon, \sqrt{\tr \dot \epsilon^2}, \phi]$. The form invariants for the stress are $G(P) = [I, \dot \epsilon]$ with corresponding coefficients $c(J) = [0,\mu (1 - \phi) J_2^{1/n - 1}]$. The damage rate is a scalar, which has only one form invariant $G_1 = 1$. The damage rate function is defined in \cref{eq:damage_rate_cr,eq:damage_rate_sf,eq:damage_rate_sh}, with $\|\dot \epsilon\| = J_2$.

\paragraph{Damage2}
This model is described by \textcite{borstad2016constitutive}.
Similarly to \cite{albrecht2014fracture}, this model includes a scalar state variable $\phi$ representing the ice damage, which advects with the ice. The inputs, scalar invariants, form invariants, and stress coefficients are the same here as for the previous Damage model.
The Damage model, however, expresses the evolution of $\phi$ as an ODE, in the form
$$
\frac{D\phi}{Dt} = s(\phi, \dot{\epsilon}),
$$
whereas the Damage2 model can express the evolution of $\phi$ as an ODE,
$$
\frac{D\phi}{Dt} = s(\phi, \dot{\epsilon}, \frac{D\dot{\epsilon}}{Dt}),
$$
or as a differential algebraic equation (DAE),
$$
s(\phi, \frac{D\phi}{Dt}, \dot{\epsilon}, \frac{D\dot{\epsilon}}{Dt}) = 0.
$$
Here $s$ is still a constitutive relation, but it is implicit. The separation of concerns used by CRIKit, where CRIKit  provides CRs that can be evaluated at points in the mesh and a simulation library models the experiment, still holds, but the simulation library must include a DAE solver, such as \cite{abhyankar2018petsc}. 
The choice here of ODE or DAE depends on how $\frac{D\dot{\epsilon}}{Dt}$ is expressed. In the steady-state infinitesimal strain-rate case, it can be explicitly computed using spatial gradients of the velocity as $\frac{D\dot{\epsilon}}{Dt} = \dot{\epsilon}(\vec{u} \cdot \nabla \vec{u})$, which allows the explicit ODE form. 
Because of the instantaneous relationship between the damage and $\dot{\epsilon}$, this approach results in an stiff ODE that requires a small time step to model accurately.
An alternative is solving for $\frac{D\phi}{Dt}$ and $\frac{D\dot{\epsilon}}{Dt}$ together as two unknown fields, which requires the implicit DAE form.

In this model, as the strain rate increases, the stress increases according to the standard power law until a threshold stress $\tau_0$ is reached. At the threshold stress, further increases in strain rate cause cause the damage to increase, such that the effective stress lies along the curve $b(J_2)$. This model includes history dependence; any accumulated damage does not change when the strain rate then decreases below the threshold strain rate.
This is described in the following equations, assuming $\phi$ is initialized to 0.
\begin{align}
a(J_2, \phi) &= (1 - \phi) \mu J_2^{1/n} \\
b(J_2) &= \tau_0 \exp\left(-\frac{J_2 - \dot \epsilon_0}{\dot \epsilon_0 (\kappa - 1)}\right)
\\
\widetilde \phi &\gets 1 - \left( \frac{J_2}{\dot \epsilon_0} \right)^{-\frac{1}{n}} \exp\left(-\frac{J_2 - \dot \epsilon_0}{\dot \epsilon_0(\kappa - 1)} \right)
\\
\phi &\gets \max(\phi, \widetilde \phi) = \begin{cases}
\widetilde \phi & \text{if $b(J_2) < a(J_2, \phi)$} \\
\text{doesn't change} & \text{else}
\end{cases}
\\
\tau &\gets (1 - \phi) \mu J_2^{1/n-1}\dot \epsilon 
\end{align}

The stress CR can be expressed in Wineman-Pipkin form identically to the previous damage model by \textcite{albrecht2014fracture}.
The damage rate can be derived by increasing the damage such that $b(J_2)$ is never less than $a(J_2, \phi)$.
\begin{align}
J_2 &\gets \|\dot \epsilon\| \\
\frac{D\widetilde \phi}{DJ_2} &\gets \left( \frac{J_2}{\dot \epsilon_0} \right)^{-\frac{1}{n}} \exp\left(-\frac{J_2 - \dot \epsilon_0}{\dot \epsilon_0(\kappa - 1)} \right) \left(\frac{1}{n} \left(\frac{J_2}{\dot \epsilon_0} \right)^{-1} + \frac{1}{\dot \epsilon_0(\kappa - 1)} \right) \\
\frac{DJ_2}{Dt}
&\gets \frac{1}{J_2} \dot \epsilon : \frac{D\dot \epsilon}{Dt}
\\ \frac{D\widetilde \phi}{Dt} &\gets \frac{D\widetilde \phi}{DJ_2}\frac{DJ_2}{Dt} \\
\frac{D\phi}{Dt} &\gets
\begin{cases}
\frac{D\widetilde \phi}{Dt} & \text{if $b(J_2) < a(J_2, \phi)$} \\
0 & \text{else}
\end{cases}
\end{align}

The derivative of $h$ with respect to $J_2$ is straightforward, but the material derivative of $J_2$ depends on the material derivative of the unknown $\vec{u}$.
To handle the $\frac{DJ_2}{Dt}$ term, the system must be expressed as a
differential-algebraic equation (DAE). I.e., instead of having an explicit
expression for the unknown $\frac{D\phi}{Dt}$, we have an equation that must be
satisfied for new unknowns $\frac{D\phi}{Dt}$ and $\frac{D\vec{u}}{Dt}$.
\begin{align}
\frac{D\phi}{Dt} &= s\left(\dot \epsilon(u), \phi, \dot \epsilon \left(\frac{D\vec{u}}{Dt}\right)\right)
\end{align}
We can define the CR to output $s$, and the resulting
implicit system can be solved for $\frac{D\phi}{Dt}$ and $\frac{D\vec{u}}{Dt}$. The function $s$ is a scalar function whose value does not depend on the choice of coordinates. Therefore, it can be expressed in Wineman-Pipkin form as a function of the scalar invariants of $P = [\dot \epsilon(u), \phi, \dot \epsilon \left(\frac{D\vec{u}}{Dt}\right)]$. The relevant invariants are $J_\phi = [\sqrt{\tr{\dot \epsilon(u)^2}}, \sqrt{\tr \dot \epsilon(u) \dot \epsilon \left(\frac{D\vec{u}}{Dt}\right)}]$
Note, the additional unknown $\frac{D\vec{u}}{Dt}$ necessitates adding the corresponding advection equation to the system of equations.
\begin{align}
0 &= - \vec{u} \cdot \nabla  \vec{u} + \frac{D\vec{u}}{Dt}
\end{align}

\paragraph{GOLF}

The GOLF model described by \textcite{gillet-chaulet_2005_anisotropic-flow-law} is orthotropic, which means it can be expressed using three
structure tensors $\bar{\vec{M}}_i$ with corresponding reference vectors
$\vec{n}_i$, such that $\bar{\vec{M}}_i = \vec{n}_i \vec{n}_i^T$. The stress can be written in terms of either $\bar{\vec{M}}_i$ or $\vec{n}_i$.
\begin{align}
\tau(\dot \epsilon, \bar{\vec{M}}_1, \bar{\vec{M}}_2, \bar{\vec{M}}_3)
&= \mu_0 \sum_{i=1}^3 \left[\mu_i\tr(\bar{\vec{M}}_i \dot \epsilon) \bar{\vec{M}}_i   + \mu_{i+3} \sym (\bar{\vec{M}}_i \dot \epsilon )_{\text{dev}}\right]
\end{align}
The stress can thus be expressed in terms of the scalar invariants $\tr(\bar{\vec{M}}_i \dot \epsilon) = \vec{n}_i^T \dot \epsilon \vec{n}_i$ (tabulated by \textcite{zheng1994theory}), and the form invariants $\bar{\vec{M}}_i$ and $\sym (\bar{\vec{M}}_i \dot \epsilon )_{\text{dev}}$. \textcite{zheng1994theory} notes that the latter form invariant group $[\sym (\bar{\vec{M}}_i \dot \epsilon )]$ is not irreducible and updates it to [$\sym (\bar{\vec{M}}_1 \dot \epsilon \bar{\vec{M}}_2 )$, $\sym (\bar{\vec{M}}_2 \dot \epsilon \bar{\vec{M}}_3 )$, $\sym (\bar{\vec{M}}_3 \dot \epsilon \bar{\vec{M}}_1 )$]. The GOLF parameters $\mu$ are computed using assumptions of a selected micro-macro model.
\todo{a sentence about how n/M depend on mu.}




\end{document}